\documentclass[amsmath,amssymb,preprint,showpacs]{revtex4}

\usepackage{graphicx}
\usepackage[usenames]{color}

\begin{document}

\title{Charge dynamics of the Co-doped BaFe$_2$As$_2$}
\author{A. Lucarelli$^1$, A. Dusza$^1$, F. Pfuner$^1$, P. Lerch$^2$, J.G. Analytis$^3$, J.-H. Chu$^3$, I.R. Fisher$^3$ and L. Degiorgi$^1$} \affiliation{$^1$Laboratorium f\"ur
Festk\"orperphysik, ETH - Z\"urich, CH-8093 Z\"urich,
Switzerland}
\affiliation{$^{2}$Swiss Light Source, Paul Scherrer Institute, CH-5232 Villigen-PSI, Switzerland}
\affiliation{$^3$Geballe Laboratory for Advanced Materials
and Department of Applied Physics, Stanford University, Stanford, California
94305-4045, USA and Stanford Institute for Materials and Energy Sciences, SLAC National Accelerator Laboratory, 2575 Sand Hill Road, Menlo Park, California 94025, U.S.A.}

\date{\today}

\begin{abstract}
We report on a thorough optical investigation over a broad spectral range and as a function of temperature of the charge dynamics in Ba(Co$_x$Fe$_{1-x}$)$_2$As$_2$ compounds for Co-doping ranging between 0 and 18\%. For the parent compound as well as for $x$=0.025 we observe the opening of a pseudogap, due to the spin-density-wave phase transition and inducing a reshuffling of spectral weight from low to high frequencies. For compounds with 0.051$\le x \le$ 0.11 we detect the superconducting gap, while at $x$=0.18 the material stays metallic at all temperatures. We describe the effective metallic contribution to the optical conductivity with two Drude terms, representing the combination of a coherent and incoherent component, and extract the respective scattering rates. We establish that the $dc$ transport properties in the normal phase are dominated by the coherent Drude term for 0$\le x \le$0.051 and by the incoherent one for 0.061$\le x \le$0.18, respectively. Finally through spectral weight arguments, we give clear-cut evidence for moderate electronic correlations for 0$\le x \le$0.061, which then crossover to values appropriate for a regime of weak interacting and nearly-free electron metals for $x\ge$0.11.

\end{abstract}



\maketitle

\section{Introduction}
The discovery of the iron-pnictides \cite{kamihara,rotter} generated a lot of interest, primarily because high-temperature superconductivity is possible in materials without CuO$_2$ planes, and also induced a frenetic search for possible common mechanisms between them and the high-temperature superconducting cuprates. Furthermore, the novel class of iron-pnictide based superconductors arose to an interesting playground, in order to study the impact of electronic correlations with respect to the emergence of structural/magnetic and superconducting phase transitions. The major difference among the phase diagram of the copper-oxide and the iron-pnictide superconductors is that the correlated metallic state in the cuprates derives through chemical doping of parent compounds in the strongly correlated Mott insulting state, while the parent compounds of the iron-pnictides are magnetically ordered bad metals. This means that magnetic-ordering only partially gaps the Fermi surface (FS) inducing a spin-density-wave (SDW) broken-symmetry ground state without initiating an insulating state. Upon doping, the structural and SDW transitions are suppressed for the benefit of superconductivity \cite{chu}.

A recent optical study on LaFePO \cite{basov} gave evidence of electronic correlations in the iron-pnictides, where the kinetic energy of the electrons is reduced to half of that predicted by band theory of nearly free electrons. This was found to imply that these systems could be on the verge of the Mott (insulator) transition. It may then be argued that transport in the iron-pnictides lies between the band-like itinerant and the Mott-like local magnetic moment extremes \cite{basov}, and that both coherent and incoherent excitations must be present. Optical investigation on several iron-pnictides ($A$Fe$_2$As$_2$, $A$=Ba, Sr, Ca, Eu and Ni) further reveals the presence of two subsystems; a coherent Fermi liquid one out of where superconductivity evolves, and a temperature independent incoherent one, acting as background to the excitation spectrum but still affected by superconductivity \cite{dressel}.

Many-body effects, as electron-electron or electron-phonon interactions and even magnetic fluctuations, have a tremendous impact on the excitation spectrum, leading to characteristic fingerprints, like the opening of energy (correlation) gaps and potentially anomalous redistribution of spectral weight, so well documented in the cuprates \cite{timusk}. Optical methods are in principle an ideal spectroscopic tool for addressing these issues. Our strategy consists here in comparing the electrodynamic response of Ba(Co$_x$Fe$_{1-x}$)$_2$As$_2$ for several Co-dopings, which belong to the so-called 122 family and are prominent examples of oxygen-free iron-pnictide superconductors. We systematically span the portion of the phase diagram ranging from the SDW state for the parent compound ($x$=0) to the proximity of the antiferromagnetic state to the superconducting phase ($x$=0.025), as well as from the superconducting compounds ($x$=0.051, 0.061 and 0.11) up to the single metallic phase ($x$=0.18). Particular emphasis will be devoted to the impact of the various transitions on the charge dynamics in these Co-doped 122 compounds. Our primary goal is to exploit their electrodynamic response in order to shed light  on the scattering processes of the itinerant charge carriers as well as to establish with spectral weight argument their degree of electronic correlations.

\section{Experiment and Results}
Single crystals of Ba(Co$_x$Fe$_{1-x}$)$_2$As$_2$ with $x$=0, 2.5\%, 5.1\%, 6.1\%, 11\% and 18\% were grown from a self flux using similar conditions to published methods \cite{chu,wang}. The crystals have a plate-like morphology, with the $c$-axis perpendicular to the plane of the plates, and grow up to several millimeters on a side. Our specimens were from the same batch used for the $dc$ transport characterization \cite{chu}. For $x=0$ the coincident structural (tetragonal-orthorhombic) and SDW transitions occur at $T_{TO}$=$T_{SDW}$=135 K, while for $x=0.025$ they develop at $T_{TO}$= 98 K and $T_{SDW}$= 92 K, respectively. The compound with $x=0.051$ undergoes first a structural transition at $T_{TO}$= 50 K and then a SDW one at $T_{SDW}$= 37 K. In the latter compound, a superconducting state finally develops below $T_c$= 19 K. At optimal doping $x=0.061$ as well as for $x$=0.11, there is only the transition into the superconducting state at $T_c$= 23 K and 14 K, respectively. The $x$=0.18 compound does not undergo any phase transitions and is metallic at all temperatures \cite{chu}.

We perform optical reflectivity measurements as a function of temperature from the far-infrared (FIR) up to the ultraviolet (UV). This is the prerequisite in order to perform reliable Kramers-Kronig (KK) transformation from where we get the phase of the complex reflectance and then all optical functions, including the real part $\sigma_1(\omega)$ of the complex optical conductivity. To this end, we extended the $R(\omega)$ spectra with the Hagen-Rubens (HR) extrapolation for $\omega\rightarrow 0$ in the metallic phase \cite{gruner,wooten} and by setting $R(\omega)$ equal to total reflection below a characteristic frequency $\omega_g$ (see below) in the superconducting state. The $\sigma_{dc}$ values used in the HR-expression coincide with $dc$ transport data collected on samples from the same batches \cite{chu}. At high frequencies the standard extrapolations $R(\omega)\sim \omega^{-s}$ with 2$\le s \le$4 were employed \cite{wooten,gruner}.

\begin{figure}[!tb]
\center
\includegraphics[width=12cm]{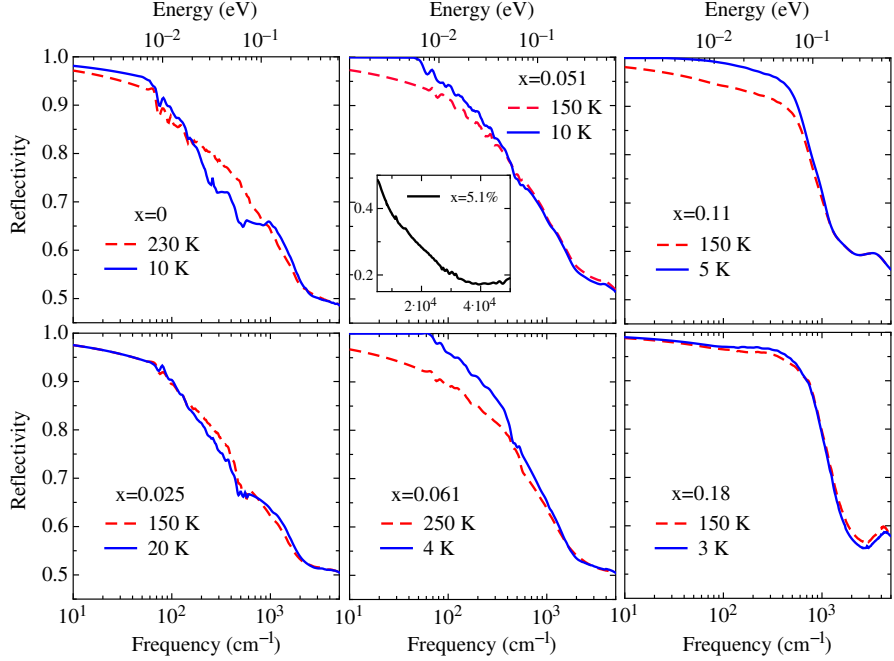}
\caption{(color online) Optical reflectivity of Ba(Co$_x$Fe$_{1-x}$)$_2$As$_2$ for $x$ ranging from 0 to 18\% in the spectral range below 5000 cm$^{-1}$ at selected temperatures above and below the various phase transitions. The inset displays $R(\omega)$ at 300 K for $x$=0.051 up to the UV spectral range, which is representative for all compounds.} \label{Refl.}
\end{figure}

Figure 1 shows the optical reflectivity $R(\omega)$ for all investigated compounds below 5000 cm$^{-1}$ at selected temperatures for the normal state as well as for the SDW or superconducting state. The inset emphasizes the high frequency $R(\omega)$ spectrum up to the UV spectral range, which is representative for all dopings. For Co-dopings $x\le$0.061, $R(\omega)$ is reminiscent of an overdamped-like behavior, displaying a gentle increase of $R(\omega)$ at frequencies from UV to the mid-infrared (inset Fig. 1) and then a steeper one below about 500 cm$^{-1}$. For Co-dopings $x\ge$0.11, a sharper plasma edge in $R(\omega)$ develops below about 2000 cm$^{-1}$, leading to larger $R(\omega)$ values than for $x\le$0.061. This is consistent with the enhanced metallicity with increasing Co-doping, as evinced from the $dc$ transport properties \cite{chu}. The temperature dependence extends over a large energy interval up to about 3000 cm$^{-1}$. This already anticipates an important reshuffling of spectral weight up to energies much larger than the energy scales defined by the phase transition critical temperatures. For $x$=0 in the SDW state, we recognize the depletion of $R(\omega)$ in the infrared range and its low-frequency enhancement at low temperatures, consistent with previous data \cite{hu,pfuner}. Such a behavior turns out to be less pronounced for the $x$=0.025 doping. For $x$=0.051, 0.061 and 0.11, $R(\omega)$ progressively increases in FIR with decreasing temperature and approaches total reflection at the finite frequency $\omega_g$ in the superconducting state (i.e., $T\ll T_c$). $R(\omega)$ for $x$=0.18 is metallic-like at all temperatures. Rather sharp and weak features at about 80 and 260 cm$^{-1}$ are identified in $R(\omega)$ of the parent compound at low temperatures, which can be ascribed to the lattice vibrational modes \cite{hu,homes,vandermarel,bernhard}. The enhanced metallicity of $R(\omega)$ for higher Co-dopings screens these modes. We do not observe any anomalies in the phonon features or the appearance of new modes when undergoing the structural phase transition \cite{phonon}. 

\begin{figure}[!tb]
\center
\includegraphics[width=12cm]{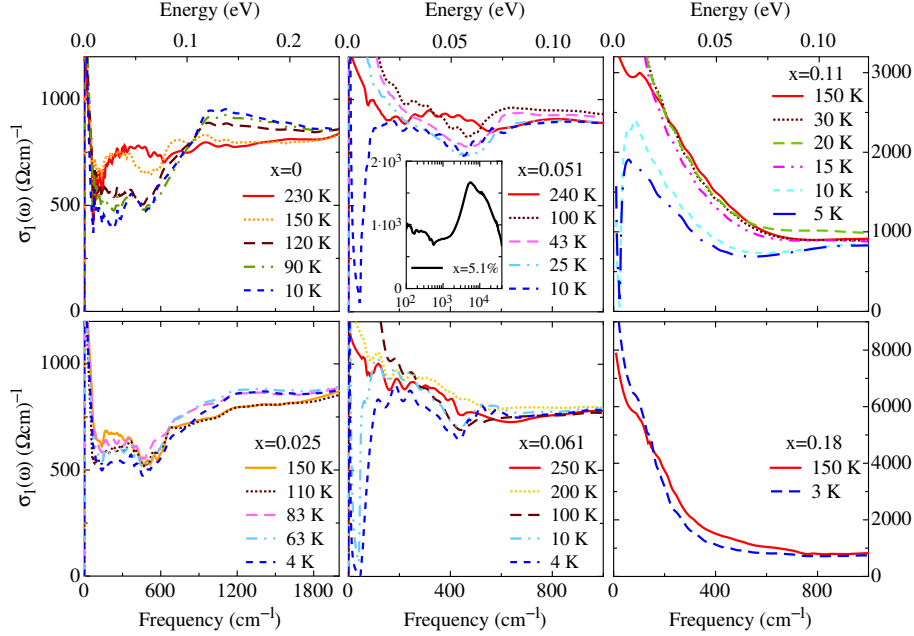}
\caption{(color online) Real part $\sigma_1(\omega)$ of the optical conductivity for Ba(Co$_x$Fe$_{1-x}$)$_2$As$_2$ with $x$ ranging from 0 to 18\% in the far and mid-infrared spectral range at selected temperatures above and below the various phase transitions. The inset displays $\sigma_1(\omega)$ at 300 K for $x$=0.051, emphasizing its representative shape for all compounds at high frequencies up to the UV.} \label{sigma}
\end{figure}

Figure 2 highlights the temperature dependence of $\sigma_1(\omega)$ for all dopings in the energy ranges pertinent to the SDW and superconducting transition. The resulting $\sigma_1(\omega\rightarrow 0)$ limits at $T>T_{SDW}$ or $T_c$ are in fair agreement with the $dc$ results, thus confirming the consistency of the KK procedure. The excitation spectrum at high frequencies, shown in the inset of Fig. 2, is identical for all Co-dopings. There is a strong absorption band peaked at about 5000 cm$^{-1}$, further characterized by a broad high frequency tail \cite{hu,vandermarel} and generally ascribed to the contribution due to the electronic interband transitions. $\sigma_1(\omega)$ for $x=0$ suddenly decreases below 800 cm$^{-1}$ at temperatures below $T_{SDW}$. This leads to a depletion in the range between 200 and 800 cm$^{-1}$ (Fig. 2), inducing a removal of spectral weight. The depletion as well as the peak at about 800 cm$^{-1}$ in $\sigma_1(\omega)$ are indicative for the opening of a pseudogap, which we identify with the SDW single particle excitation \cite{hu,pfuner}. For $x$=0.025 Co-doping, the depletion as well as the (pseudo)gap feature in $\sigma_1(\omega)$ are less evident and pronounced, even though there is a spectral weight redistribution, leading again to its overshoot above 700 cm$^{-1}$ for temperature below $T_{SDW}$. The signatures for the SDW pseudogap-like excitation as well as the related spectral weight redistribution are no longer well distinct for the $x$=0.051 Co-doping \cite{comment51}. For this latter compound as well as at and above the optimal Co-doping ($x=$0.061 and 0.11, respectively), the total reflection at $\omega\le\omega_g$ for $T\ll T_c$ leads instead to the opening of the superconducting gap \cite{vandermarel,bernhard,dressel,dressel2}. The removed spectral weight is shifted into the collective excitation at zero frequency. While our data in the superconducting state overall agree with previous work, they do not extend low enough in frequencies, in order to allow an elaborated analysis in terms of multiple superconducting gaps \cite{vandermarel,bernhard}. Finally, the electrodynamic response of the $x$=0.18 compound merely displays a simple metallic behavior. 

Overall, there are three energy intervals characterizing $\sigma_1(\omega)$ for all Co-dopings; the effective metallic contribution at low frequencies, the mid-infrared (MIR) band covering the energy range between 500 and 1500 cm$^{-1}$ and the electronic interband transitions with onset at about 2000 cm$^{-1}$ and peaked at 5000 cm$^{-1}$. The metallic part as well as the MIR band turn out to experience the strongest temperature dependence at $T_c$ and/or $T_{SDW}$ (Fig. 2), while the high frequency excitations (inset Fig. 2) are temperature independent. 

\begin{figure}[!tb]
\center
\includegraphics[width=12cm]{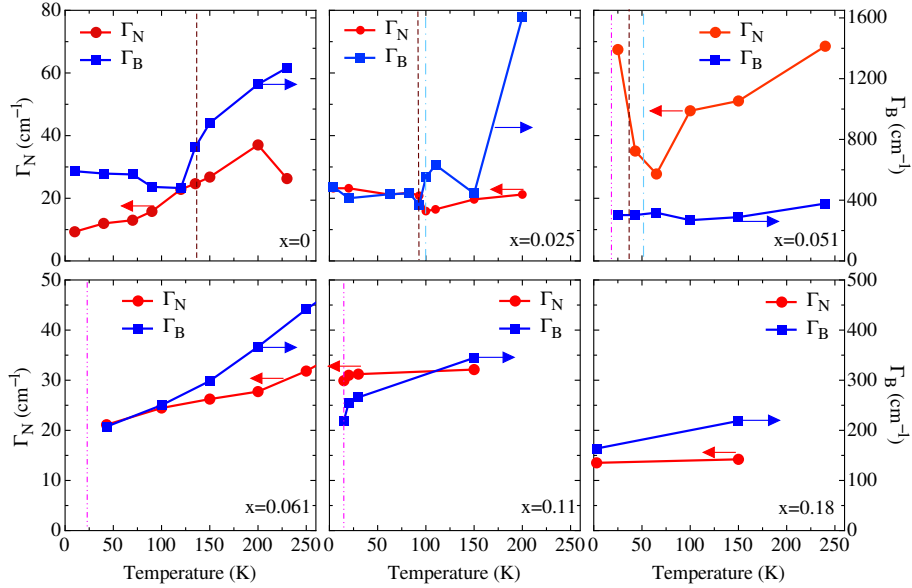}
\caption{(color online) Temperature dependence of the scattering rate of the narrow ($\Gamma_N$) and broad ($\Gamma_B$) Drude terms (eq. (1)), used to fit the effective metallic contribution to the optical conductivity of Ba(Co$_x$Fe$_{1-x}$)$_2$As$_2$ with $x$ ranging from 0 to 18\%. The thin dashed-dotted, dashed and dashed-dotted-dotted vertical lines mark, when appropriate, the structural (tetragonal-orthorhombic), SDW and superconducting transitions, respectively.} \label{scattering}
\end{figure}

\section{Discussion}
Since the superconducting properties were intensively studied by different groups on selected members of the iron-pnictides family \cite{dressel,vandermarel,bernhard,dressel2}, we will mainly address here the normal and SDW state properties. To this end, we apply the common and well established phenomenological Drude-Lorentz approach \cite{wooten,gruner}. In order to account for the various bands crossing the Fermi level \cite{singh,ARPES} and consistent with previous investigations \cite{dressel}, we ascribe two Drude contributions (one narrow and one broad) to the effective metallic part of $\sigma_1(\omega)$ and a series of Lorentz harmonic oscillators (h.o.) for all excitations (phononic and electronic) at finite frequencies: 
{\setlength\arraycolsep{2pt}
\begin{eqnarray}
\nonumber \tilde{\epsilon}(\omega) & = & \epsilon_1(\omega)
+i\epsilon_2(\omega) =
\\ & = & \epsilon_{\infty}-\frac{\omega_{PN}^2}{\omega^2-i \omega
\Gamma_N}-\frac{\omega_{PB}^2}{\omega^2-i \omega
\Gamma_B}+\sum_j \frac{S_j^2}{\omega_j^2-\omega^2-i \omega
\gamma_j},
\end{eqnarray}}where $\epsilon_{\infty}$ is the optical dielectric constant,
$\omega_{PN/B}$ and $\Gamma_{N/B}$ are the plasma frequencies and the widths
of the narrow and broad Drude peaks, whereas $\omega_j$, $\gamma_j$, and $S^2_j$ are
the center-peak frequency, the width, and the mode strength for
the $j$-th Lorentz harmonic oscillator (h.o.), respectively. $\sigma_1(\omega)$ is then obtained from $\sigma_1(\omega)=\omega
\epsilon_2(\omega)/4\pi$. The approach with two Drude terms implies the existence of two electronic subsystems. The narrow Drude term is identified with the coherent part of the excitation spectrum, while the broad one acts as background to $\sigma_1(\omega)$ and represents its incoherent contribution. Besides the sharp and weak h.o.'s for the phonon modes, three broad ones for the strong absorption features leading to the peak at about 5000 cm$^{-1}$ and one for the MIR band were additionally considered in our fits \cite{commentho}. The inset of Fig. 6 depicts the Drude terms as well as the h.o. for the MIR band, which fully account for the temperature dependence of the spectra. We systematically apply this fit procedure for each Co-dopings at all temperatures, obtaining everywhere a comparable good fit quality (e.g., the spectra at 120 K for $x$=0, inset Fig. 6). 

\begin{figure}[!tb]
\center
\includegraphics[width=12cm]{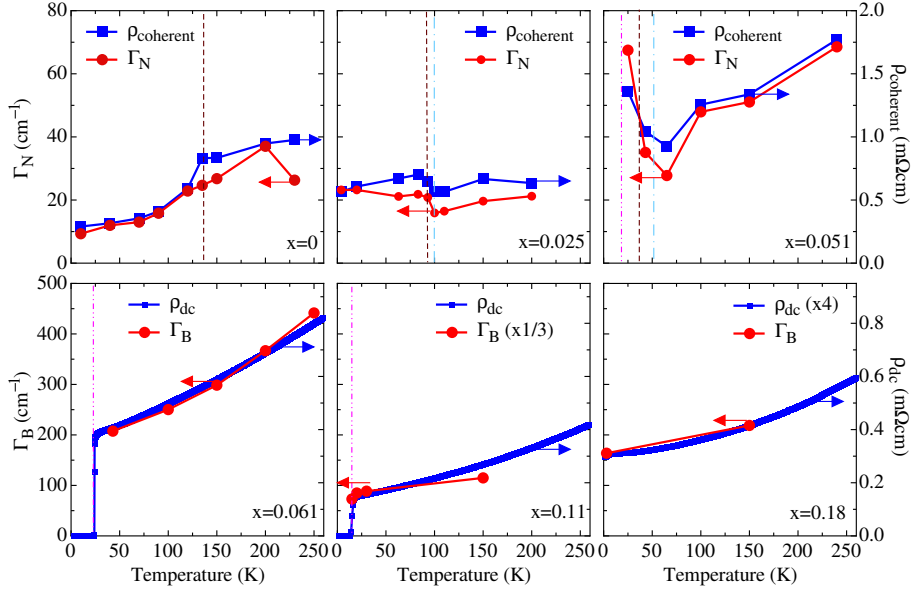}
\caption{(color online) Upper panels: comparison between the scattering rate of the narrow Drude term ($\Gamma_N$) and the coherent contribution to the $dc$ transport properties (eq. (2)) for Ba(Co$_x$Fe$_{1-x}$)$_2$As$_2$ with $x$ ranging from 0 to 5.1\%. Lower panels: comparison between the scattering rate of the broad Drude term ($\Gamma_B$) and the $dc$ transport properties ($\rho_{dc}(T)$, Ref. \onlinecite{chu}) for Ba(Co$_x$Fe$_{1-x}$)$_2$As$_2$ with $x$ ranging from 6.1 to 18\%. The thin dashed-dotted, dashed and dashed-dotted-dotted vertical lines mark, when appropriate, the structural (tetragonal-orthorhombic), SDW and superconducting transitions, respectively.} \label{transport}
\end{figure}

First, we extract the scattering rate $\Gamma_N$ and $\Gamma_B$ of both narrow and broad Drude term, respectively. $\Gamma_N$ and $\Gamma_B$ are shown in Fig. 3. At the phase transition temperatures for 0$\le x \le$0.051 both scattering rates display anomalies, while for 0.061$\le x\le$0.18 they monotonically decrease or stay constant with decreasing temperatures, consistent with a rather conventional metallic behavior. More compelling at this point is the comparison with the $dc$ transport properties. Inspired by Wu \emph{et al.} (Ref. \onlinecite{dressel}), we assume that $\sigma_{dc}(T)$ can be expressed as the $\omega\rightarrow$0 limit of $\sigma_1(\omega)$ for both narrow and broad Drude term. Therefore, we argue that the coherent contribution to the $dc$ transport is given by:
{\setlength\arraycolsep{2pt}
\begin{eqnarray}
\rho_{dc}^{coherent}=(\rho_{dc}^{-1}(T)-\sigma_{dc}^{broad Drude})^{-1},
\end{eqnarray}}
where $\rho_{dc}(T)$ is the measured resistivity from Chu \emph{et al.} (Ref. \onlinecite{chu}) and $\sigma_{dc}^{broad Drude}$ is the $\omega\rightarrow$0 limit of the broad Drude term. As shown in Fig. 4 (upper panels), we discover that $\Gamma_N$ scales to the calculated quantity $\rho_{dc}^{coherent}(T)$ for Co-dopings $x\le$0.051. Interestingly enough, $\Gamma_N$ and $\rho_{dc}^{coherent}(T)$ for $x$=0 follows a $T^2$ dependence at low temperatures, consistent with previous findings and uncovering a Fermi liquid behavior \cite{dressel}. For Co-dopings $x\ge$0.061, it turns out that $\Gamma_B$ scales fairly well with the total resistivity $\rho_{dc}(T)$ \cite{chu}, as shown in the lower panels of Fig. 4. This derives from the fact that the two Drude terms, globally describing the low frequency part of $\sigma_1(\omega)$, are no longer so well distinct for $x\ge$0.061 as for $x\le$0.051 and the incoherent contribution largely dominates $\sigma_1(\omega)$ over the coherent one. Consequently, the decomposition of the metallic contribution in $\sigma_1(\omega)$ in two independent conduction channels is not unique anymore \cite{matthiessen}. 
We thus suggest that for Co-dopings, where the FS may be gapped by the SDW instability, $\Gamma_N$ shapes the coherent contribution to the $dc$ transport properties, while for Co-dopings above $x$=0.061 the scattering rate of the broad (incoherent) Drude term mainly determines the $dc$ transport.

\begin{figure}[!tb]
\center
\includegraphics[width=12cm]{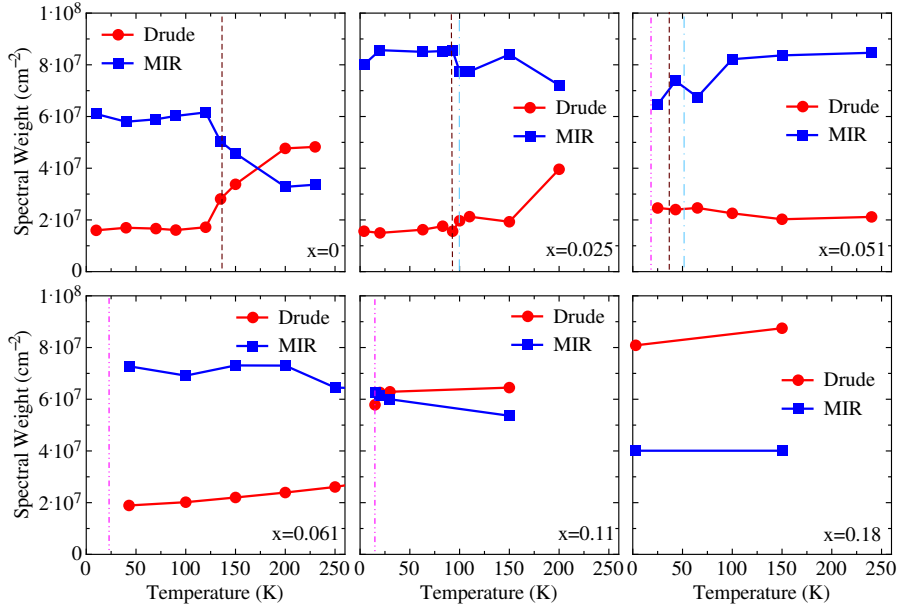}
\caption{(color online) Temperature dependence of the integrated spectral weight (eq. (3) and Ref. \onlinecite{commentweight}) of the Drude components and of the MIR band in $\sigma_1(\omega)$ (see inset Fig. 6) for Ba(Co$_x$Fe$_{1-x}$)$_2$As$_2$ with $x$ ranging from 0 to 18\%. The thin dashed-dotted, dashed and dashed-dotted-dotted vertical lines mark, when appropriate, the structural (tetragonal-orthorhombic), SDW and superconducting transitions, respectively.} \label{spectral weight}
\end{figure}

Another relevant quantity, characterizing the electrodynamic response, is the integrated spectral weight encountered in $\sigma_1(\omega)$ up to a cut-off frequency $\omega'$, which is achieved through the well-known $f$-sum rule \cite{wooten,gruner}:
{\setlength\arraycolsep{2pt}
\begin{eqnarray}
SW(\omega')=\frac{120}{\pi}\int_{0}^{\omega'}\sigma_1(\omega)d\omega.
\end{eqnarray}}It turns out, that the sum rule is almost fully satisfied between 2000 and 3000 cm$^{-1}$ for all Co-dopings, irrespective whether there is a phase transition into a SDW or a superconducting state. This is at variance to the high-temperature superconducting cuprates, where the spectral weight, merging into the collective state at $\omega$=0 as consequence of the opening of the superconducting gap along the $c$-axis orthogonal to the CuO$_2$ plane, is actually transferred from very high energy scales. This means equivalently that in the cuprates only 50\% of the weight pertaining to the collective state is recovered at mid-infrared energies \cite{basovweight}.

\begin{figure}
\center
\includegraphics[width=12cm]{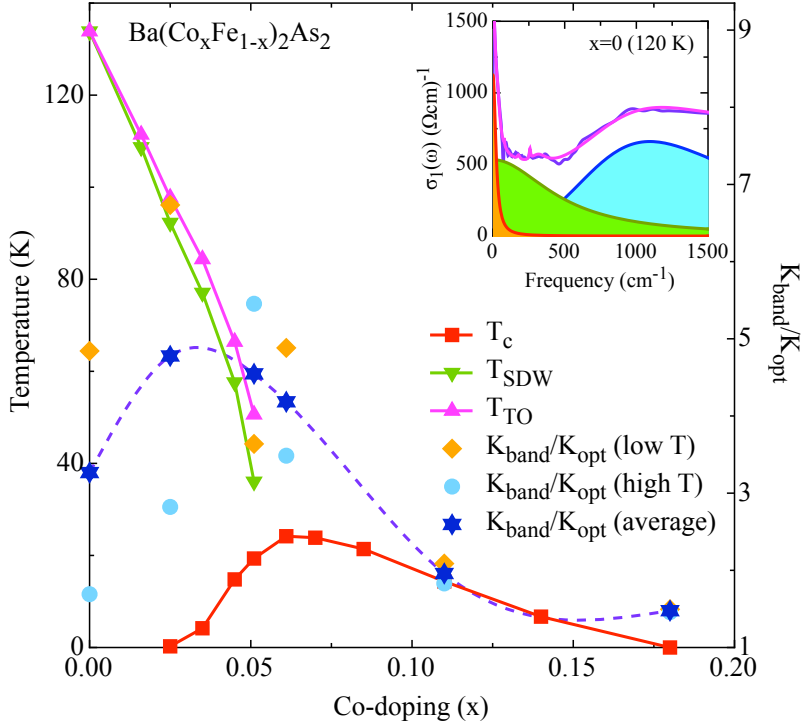}
\caption{(color online) Phase diagram of Ba(Co$_x$Fe$_{1-x}$)$_2$As$_2$, reproduced from Ref. \onlinecite{chu} (left y-axis), and the ratio $K_{band}/K_{opt}$ (eq. (4)) calculated at low and high temperatures as well as the average of both (right y-axis). All data-interpolation are spline lines as guide to the eyes. The inset displays the optical conductivity at 120 K for the parent compound ($x$=0) with the total Drude-Lorentz fit and its low frequency components; i.e., the narrow and broad Drude terms as well as the mid-infrared (MIR) h.o. (see text and eq. (1)). The shaded areas emphasize the respective spectral weights ($\int\sigma_{1N}(\omega)d\omega, \int\sigma_{1B}(\omega)d\omega$ and $\int\sigma_1^{MIR}(\omega)d\omega$, eq. (4)).} \label{phase diagram}
\end{figure}

In order to investigate how the spectral weight redistributes as a function of temperature among the various components in $\sigma_1(\omega)$, we can further exploit our phenomenological approach. Figure 5 summarizes indeed the temperature dependence of the spectral weight encountered in the Drude terms and in the MIR band (inset of Fig. 6). For the parent compound ($x$=0) there is an obvious depletion of weight in the Drude part at $T_{SDW}$, which is reshuffled into the MIR band. At $x$=0.025 there is a similar yet more modest redistribution of weight at $T_{SDW}$. At $x$=0.051 the Drude part of $\sigma_1(\omega)$ slightly gains weight with decreasing temperature at the cost of the MIR part \cite{comment}. We tend to believe that upon Co-doping disorder is induced in the system such that the favorable nesting conditions, so crucial for magnetic order, are progressively suppressed. Disorder may then favor the washing out of the SDW gap feature, as observed for $x=0.025$ and 0.051. We should also add, that for the latter doping $T_c$ is pretty close to $T_{SDW}$. Therefore, by the time the temperature is low enough to allow a fully developed SDW (pseudo)gap, the system is already in the superconducting state and it is no surprise that almost no signatures of the SDW transition are optically detected for $x=0.051$ \cite{comment51}. For higher Co-dopings (i.e., $x\ge$0.061), the spectral weight of the Drude and MIR components are almost temperature independent and interestingly enough the Drude weight increases with respect to the MIR one, the latter observation being an indication for the enhanced metallicity with increasing Co-dopings.

We propose a scenario where the conduction band derives from $d$-states and splits into two parts: a purely itinerant one close to the Fermi level and represented by the two Drude components as well as by a bottom part with states below the mobility edge and thus rather localized. This latter part gives rise to the MIR band in $\sigma_1(\omega)$, which turns out to be strongly temperature dependent upon magnetic ordering and affected by the opening of the SDW gap. We can then consider the following ratio:
{\setlength\arraycolsep{2pt}
\begin{eqnarray}
K_{opt}/K_{band}=\frac{\int\sigma_{1N}(\omega)d\omega+\int\sigma_{1B}(\omega)d\omega}{\int\sigma_{1N}(\omega)d\omega+\int\sigma_{1B}(\omega)d\omega+\int\sigma_1^{MIR}(\omega)d\omega},
\end{eqnarray}}where $\sigma_{1N}(\omega)$, $\sigma_{1B}(\omega)$ and $\sigma_1^{MIR}(\omega)$ are the components of the optical conductivity due to the narrow and broad Drude terms, and to the MIR band within our phenomenological approach (eq. (1) and inset of Fig. 6), respectively. Equation (4) represents the ratio between the spectral weight encountered in $\sigma_1(\omega)$ in the Drude components ($K_{opt}$) and the total spectral weight collected in $\sigma_1(\omega)$ up to the onset of the electronic interband transitions (i.e., Drude components together with the MIR absorption feature, $K_{band}$). Such a ratio is an alternative estimation, exclusively from the experimental findings, of the ratio between the optical kinetic energy extracted from $\sigma_1(\omega)$ and the band kinetic energy extracted from the band structure within the tight-binding approach \cite{basov}. Our value of  $K_{opt}/K_{band}$ for $x$=0 is consistent with the previous estimation \cite{basov}, therefore reinforcing the validity of our alternative method. The inverse of $K_{opt}/K_{band}$, which then defines the degree of electronic correlations, is plotted in Fig. 6 within the phase diagram of the Co-doped 122 iron-pnictides \cite{chu}. $K_{band}/K_{opt}$ thus tracks the evolution of the superconducting dome in the phase diagram of Ba(Co$_x$Fe$_{1-x}$)$_2$As$_2$. Interestingly enough, electronic correlations seem to be stronger for the parent-compound and for Co-dopings $x\le$0.061 than for those in the overdoped range. There is indeed evidence for a crossover from a regime of moderate correlations for $x\le$0.061 to a nearly free and non-interacting electron gas system for $x\ge$0.11. 

\section{Conclusions}
We provided a comprehensive optical investigation of Ba(Co$_x$Fe$_{1-x}$)$_2$As$_2$ with several Co-dopings, scanning the complete phase diagram. Besides establishing a direct relationship between the scattering rates and the $dc$ transport properties, we shed light on the spectral weight redistribution between the itinerant and the localized portion of the conduction band. We argue that the 122 iron-pnictides fall in the regime of moderate electronic correlations for the parent compound and close to optimal doping, while a conventional metallic behavior in the nearly free electron limit is recovered at the opposite end of the superconducting dome in their phase diagram.\\

\emph{Note added - }While preparing this manuscript for submission, we came aware of the work by Nakajima et al. \cite{nakajima}, which presents similar data for part of the samples discussed here.

\begin{acknowledgments}
The authors wish to thank M. Dressel, L. Benfatto, D. Basov, M. Qazilbash, R. Hackl and A. Chubukov for fruitful discussions, and P. Butti for valuable help in collecting part of the data. This work has been supported by the Swiss National Foundation for the Scientific Research
within the NCCR MaNEP pool. This work is also supported by the
Department of Energy, Office of Basic Energy Sciences under
contract DE-AC02-76SF00515.
\end{acknowledgments}


\begin{thebibliography}{99}
\bibitem{phonon} The phonon features in our spectra are too weak, so that possible mode splitting, due to the lowering of the crystal symmetry, as well as signature for mode softening are beyond our experimental resolution.
\bibitem{comment51} An attentive look to the data reveals nevertheless an incipient depletion at 100 K, which gets stronger at 43 K, a temperature between $T_{TO}$ and $T_{SDW}$. The depletion at 43 K is accompanied by a reshuffling of the spectral weight at high frequencies, resulting in the formation of a barely observable pseudogap-like feature. By approaching the superconductivity phase transition, these weak features are totally washed out. The spectral weight redistribution involved at temperatures between 100 and 10 K in the energy interval between 200 and 1000 cm$^{-1}$ is moreover very small to allow more robust considerations about possible coexistence of the SDW and superconducting state.
\bibitem{commentho} For $x$=0 and $x$=0.025, two additional h.o.'s, located in the high and low frequency tail of the narrow Drude and MIR band h.o., respectively, were also employed in order to better reproduce the shape of $\sigma_1(\omega)$. These contributions were consistently included in the spectral weight analysis of the Drude terms and MIR band in relation to Fig. 5 and 6.
\bibitem{commentweight} Equation (3) is equal to the sum of the square of the plasma frequencies for the Drude terms and to the mode strength of the h.o. for the MIR band \cite{commentho}.
\bibitem{comment} With decreasing temperature, the h.o. for the MIR absorption feature in $x$=0.051 tends to redistribute its spectral weight towards the low frequency tail of the electronic interband transitions.
\end{thebibliography}
\end{document}